\documentstyle[preprint,tighten,aps,epsfig]{revtex}

\def\lapprox{\mathrel{\mathop  {\hbox{\lower0.5ex\hbox{$\sim$}
\kern-0.8em\lower-0.7ex\hbox{$<$}}}}}
\def\gapprox{\mathrel{\mathop  {\hbox{\lower0.5ex\hbox{$\sim$}
\kern-0.8em\lower-0.7ex\hbox{$>$}}}}}
\begin{document}
%
\preprint{\vbox{\noindent
\hfill INFNFE-01-03 \\
\null \hfill MPIK Preprint 2003-001} }
\title{KamLAND, terrestrial heat sources and neutrino oscillations}

\author{G.~Fiorentini$^{(1,2)}$, T. Lasserre$^{(3)}$, M.~Lissia$^{(4,5)}$, 
        B.~Ricci$^{(1,2)}$ and S.~Sch\"onert$^{(6)}$}

\address{
$^{(1)}$Dipartimento di Fisica dell'Universit\`a di Ferrara, I-44100
Ferrara, Italy\\
$^{(2)}$Istituto Nazionale di Fisica Nucleare, Sezione di Ferrara, 
I-44100 Ferrara, Italy \\
$^{(3)}$DSM/DAPNIA/SPP, CEA/Saclay
F-91191 Gif-Sur-Yvette CEDEX, France \\
$^{(4)}$Istituto Nazionale di Fisica Nucleare, Sezione di Cagliari
I-09042 Cagliari, Italy \\
$^{(5)}$Dipartimento di Fisica dell'Universit\`a di Cagliari, I-09042
Cagliari, Italy\\
$^{(6)}$Max-Planck-Institut f\"ur Kernphysik Heidelberg, Postfach 103980,
D-69029 Heidelberg, Germany
}
\date{\today}
\maketitle
\begin{abstract}
We comment on the first indication of geo-neutrino events from KamLAND
and on the prospects for understanding Earth energetics. Practically all
models of terrestrial heat production are consistent with data within the
presently limited statistics, the fully radiogenic model being closer to
the observed value ($\approx 9$ geo-events).
In a few years KamLAND should collect sufficient data for a clear
evidence of geo-neutrinos, however discrimination among models requires
a detector with the class and size of KamLAND far away from nuclear
reactors. We also remark that the event ratio from Thorium and Uranium
decay chains is well fixed $N(Th)/N(U) \simeq 0.25$, a constraint that
can be useful for determining neutrino oscillation parameters.
We show that a full spectral analysis, including this constraint, further reduces
the oscillation parameter space compared to an analysis with an energy threshold
$E_{vis}>2.6 \, MeV$. 
\end{abstract}
%
\section{Introduction}
Recently KamLAND  presented the first results \cite{kamland} on 
the search for  oscillation  of  $\bar{\nu}_e$ emitted from distant
power reactors. 
Electron antineutrinos are detected by means of inverse beta decay,
\begin{equation}
\label{eqdecay}
\bar{\nu}_e+ p \rightarrow e^+ + n - 1.80 \, MeV \quad ,
\end{equation}
by looking at the prompt energy deposited by the positron 
($E_{vis}=2m_e+ E_{kin}$, where the kinetic energy of the positron is 
$E_{kin}=E_{\bar{\nu}}-1.80\, MeV$ )
accompanied by the signal of the neutron from $n+p \rightarrow d+\gamma$.
With an exposure of 162 ton$\cdot$yr a clear deficit has been observed,
however various combinations of oscillation parameters 
describe well the shape of
the positron spectrum. The best fit value reported in \cite{kamland}, including
the geo-neutrino fluxes as free parameters, corresponds to
$\sin ^2 2\theta\cong 0.91$ and  $\Delta m^2\cong 6.9 \cdot 10^{-5}\,eV^2$ .

Terrestrial antineutrinos, emitted by the $\beta$-decay of the progenies
of  $^{238}U$ and $^{232}Th$ in the Earth's 
interior, contribute to the low energy part of the detected signal, the
maximal
$E_{vis}$ being 2.48  and 1.46 MeV respectively.
From a fit to the experimental data the KamLAND collaboration reported 4 events 
associated to $^{238}U$  and 5 to $^{232}Th$ \cite{kamland}. These numbers
provide a  direct insight on the radiogenic component  of the
terrestrial heat.
In this paper we comment on the implications of this first result and on
the prospects which it discloses for understanding the energetics of the Earth.
We also discuss the constraints provided by geo-neutrinos for precise 
determinations of the neutrino oscillation parameters.

\section{KamLAND and terrestrial heat sources}

Given the cross section of  (\ref{eqdecay}) and the antineutrino spectra  one can immediately
 derive the relationship between the number of events and the antineutrino fluxes (see Appendix):
\begin{eqnarray}
\label{eqNU}
N(U)  &=&13.2 \cdot  P \cdot \epsilon(U) \cdot  \Phi(U)\\
\label{eqNTh}
N(Th)  &=& 4.0  \cdot  P \cdot \epsilon(Th)  \cdot  \Phi(Th) \quad ,
\end{eqnarray}
where event numbers $N$ are calculated for an exposure of $10^{32}$ 
protons$\cdot$yr, $P$ is  the averaged survival probability and  
$\epsilon$ are the detection efficiencies
(from the values quoted in \cite{kamland}, 
we get $P=1-1/2 \sin^2 2\theta \cong 0.55$ and
 $\epsilon \cong 78.3\%$ for both $U$ and $Th$).
$\Phi(X)$ are the \underline{produced} fluxes in units of 
$10^6 \, cm^{-2} \, s^{-1}$, i.e. the fluxes which one should
observe in the absence of oscillation:
\begin{equation}
\label{eqflux}
\Phi(X)= \int _{V_{\oplus}} d^3r \frac{ \rho(\vec{r}) } {4\pi|\vec{R}-\vec{r}|^2} \,
				      \frac{C_X(\vec{r})n_X}{\tau_X m_X}\quad ,
\end{equation}
where $\vec{R}$ is the detector location,
$\rho$ is the density, $C_X$, $\tau_X$ and $m_X$ are the concentration, lifetime
and atomic mass of element $X$  and $n_X$ is the number of antineutrinos emitted per decay chain. 
The integration is performed over the Earth volume $V_{\oplus}$.

The radiogenic contribution to the  terrestrial heat is not quantitatively understood.
In ref. \cite{noi} three representative models have been considered:

a) a  naive chondritic model, where one assumes for the Earth
mass ratios typical of carbonaceous chondrites \cite{LB}:  
$[Th]/[U]= 3.8$, $[K]/[U]=7\cdot 10^4$ and $[U]/[Si] = 7.3\cdot10^{-8}$.
In this model the radiogenic heat production rate is about $30 \,TW$, 
originating mainly from $K$ decays.

b) The  Bulk Silicate Earth (BSE) model, which provides a description of geological evidence 
coherent with geochemical information  and accounts for  a radiogenic production of about  
$20 \, TW$. In this model one has: $[Th]/[U]=3.8$, $[K]/[U]=10^4$ and $[U]/[Si]= 9.4\cdot10^{-8}$.

c) A fully radiogenic model, where the abundances of $Th,\, U$ and $K$ are 
rescaled with respect to 
b) so as to account for the full terrestrial heat flow of $40\,TW$.

Uranium mass in the crust $M_{crust}(U)= 0.4 \cdot 10^{17} \,kg$ has been fixed. 
By taking 
$ M(Si)/M_\oplus = 0.15$  all other masses in the crust and in the mantle 
are obtained from the above ratios.  Uniform distributions within 
the crust and the mantle are assumed.
For each model, from the fluxes of ref. \cite{noi}
and eqs. (\ref{eqNU},\ref{eqNTh})  we obtain the number of events 
expected in the first exposure of KamLAND, see Table \ref{tabevents}.

In view of the limited statistics, it is useful to consider the sum
 of terrestrial events $N(U+Th)$.
The measured value is essentially obtained from a total of $C=32$ counts with 
$E_{vis}<2.6 \,MeV$, after subtraction of reactor events $R$ and 
background $B$: $N(U+Th)=C-R-B$.
The statistical fluctuation is thus of order $\sqrt{C}=5.7$.
Within this uncertainty all models are consistent with data, the fully 
radiogenic model being closer to the value reported in \cite{kamland}.

We remark that the ratio $N(Th)/N(U)$ is a significant indicator. In fact, the separate 
event numbers depend on the amount of radioactive materials,  on their distribution inside 
the Earth, on the antineutrinos survival probability and on the detection efficiency. On the other hand, if one assumes an approximately  uniform mass ratio $[Th]/[U]$ 
inside the Earth, the event ratio does not depend on material distribution
and 
on the survival probability, as it is clear from eq. (\ref{eqapp}). 
Assuming $\epsilon(Th)=\epsilon(U)$,
one has 
\begin{equation}
\label{eqrappo}
N(Th)/N(U)=0.065 \, [Th]/[U] \quad .
\end{equation}

For most models of terrestrial composition one has 
$[Th]/[U] \cong 3.8$, giving 
\begin{equation}
  N(Th)/N(U)\cong 0.25 \quad .
\end{equation}


On the other hand, by considering both $N(Th)$ and $N(U)$ as independent parameters
KamLAND obtains $N(Th)/N(U) \approx 1$. If confirmed with higher statistics,
this would imply $[Th]/[U] \approx 16 $, quite an unexpected value. 
However, a model with  $[Th]/[U]=16$, $[K]/[U]=10^4$ and 
$M_{crust+mantle}(U)= 0.8\cdot 10^{17} \, kg$ would provide 
the full observed heat flow, the main source 
 being $Th$  at 28 TW. The model predicts  about 5 events in KamLAND, 
half from $Th$ and half from $U$ decays. 

In order to discuss the achievable  improvements, we collect in
Table~\ref{tabconf} 
the predictions of several models, normalized to $10^{32}$ protons$\cdot$yr
~\footnote{KamLAND will presumably obtain this exposure within two 
years. We remind that KamLAND present fiducial mass is 408 tons, out of  the 
total of 1000 tons of mineral oil.}) and assuming 100\% detection efficiency.
We present the results following from \cite{noi}, together with predictions obtained
from the fluxes estimated in \cite{Chen} and some estimates from \cite{Ragh}, rescaled 
for a 0.55 survival probability (model IIb of \cite{Ragh} assumes that heat
production is fully sustained by $U$ and $Th$, omitting any contribution
from
Potassium).
From the various models, one estimates 
 $Th+U$  events  in the range 24-83 respectively.

At a site with negligible reactor flux $N(Th+U)$ could thus be measured 
with an accuracy of about 20-10\% and the different models could 
be clearly discriminated.
On the other hand, by rescaling the present KamLAND data, one expects that 
counts with $E_{vis}<2.6 \, MeV$ will be 
$C^{'} \simeq C \cdot 7.19/0.78=295$. This implies statistical 
fluctuations of about $\pm 17$ events, which possibly will allow for a 
clear evidence of geo-neutrinos, however they are too large for model 
discrimination.

All this calls for a detector with the class and the size of KamLAND, 
far away from nuclear reactors. We note that BOREXINO \cite{borexino}
will provide additional and complementary information in the future. 
Its target mass is about 300 tons and the reactor background corresponds 
to about 7 events per year below $2.6 \, MeV$, thus providing a better
signal to background ratio
\cite{Ragh,Chen}.

\section{Geo-neutrinos and oscillation parameters}

When KamLAND data at $E_{vis}\geq 2.6 MeV$ are combined with solar and
Chooz data,  the solution to the solar neutrino problems is basically
split in two near regions, called LMA-I and LMA-II \cite{Fogli,Maltoni,Bah,Nunokawa,Aliani,Holanda,Barger}.  
The  first region contains the global best fit point,
corresponding to $(\Delta m^2/10^{-5}\,eV^2, \sin^2\theta)= (7.3, 0.315)$,
whereas the second one is centered around $(15.4, 0.300)$ \cite{Fogli}.
A relevant question is thus if geo-neutrinos can be of some help for
discriminating between the two solutions.

As previously remarked, although  the total amounts of  $U$ and $Th$ inside
Earth are not well determined,  the ratio of  their abundances is rather
constrained. Estimates  for the solar system  yield $[Th]/[U]=(3.7-3.9)$
\cite{Anders}, estimates for the primitive mantle are in the range
$(3.6-4)$,  measurements of the upper continental crust give $(3.8-4.2)$,
estimates
of the bulk continental crust are in the range $(3.8-5)$ \cite{Wede,Ahrens}. By
assuming $[Th]/[U]=3.8 \pm 0.7$, from eq. (\ref{eqrappo}) we get  for the ratio of
geo-events:
\begin{equation}
\label{rapporti}
r= N(Th)/N(U)=0.25 \pm 0.05 \quad .
\end{equation}

We remark that this constraint, which has been derived by assuming an
approximately uniform distribution of $[Th]/[U]$ and  equal (distance
averaged) survival probabibilities, has actually a larger validity.

Concerning the effect of regional $[Th]/[U]$ variations, from
\cite{Chen} we derive  that $r$ is changed by less than 2\% when the
detector is placed at Kamioka, or Gran Sasso, or Tibet (on the top of  a
very thick continental crust) or at the Hawaii (sitting on the thin,  $U$-
and $Th$-poor oceanic crust). Coming to the effect of local variations,
by assuming  that within 100 $km$ from the detector the Uranium abundance
is double, [Th]/[U]=2 , one gets $r=0.22$, whereas if its is halved,
$[Th]/[U]=8$, one finds $r=0.28$. Neutrino oscillations clearly do not
affect eq. (\ref{rapporti}) if the oscillation lenghts for both $U$ and $Th$ neutrinos
are both very short or very long in comparison with some typical Earth
dimension. We have checked that the effect of finite oscillation lengths
does not change $r$ by more than 2
In conclusion, all these effects are well within the estimated 20\%
uncertainty on $r$.

\subsection{A sum rule}

In order to see the implications of this constraint,
let us first divide the KamLAND signal below $2.6 \,MeV$ in two regions:
a) $0.9<E_{vis}(MeV)<1.75$ corresponding to the first two bins of
\cite{kamland} and 
b) $1.75 <E_{vis}(MeV)<2.60$, corresponding to the next two bins.

All $Th$-events are contained in region a), whereas  a fraction $s$ of $U$
events are  in a) and in $(1-s)$ are in b). We find  $s=0.6$. The number
of geo-events $G_{a,b}$ in each region is thus:
\begin{equation}
\label{geoevent}
G_a= (r+s)N(U) \quad ; \quad G_b= (1-s) N(U) \quad .
\end{equation}
By eliminating $N(U)$ from the two equations one obtains a sum rule:
\begin{equation}
\label{sum}
(1-s)G_a- (r +s) G_b=0 \quad .
\end{equation}

For each solution, we can extract $G_{a,b}$ from KamLAND counts
$C_{a,b}$, after subtracting the estimated background  $B_{a,b}$ and
the predicted reactor events $R_{a,b}$. One can then build the quantity
$S= (1-s)G_a- (r +s) G_b$ and check if it is consistent with zero. We
remark that  $ C_{a,b}$ are essentially  independent observables and the
statistical fluctuations are not correlated. This procedure is shown in
Table \ref{table3} for the best fit points of LMA-I and LMA-II~\footnote{
As a general consideration,  we use here the values obtained from global 
analysis \cite{Fogli} and omit uncertainties related to theoretical predictions.}. 
The resulting
values of $S$, calculated for $r=0.25$ and $s=0.6$,
\begin{equation}
\label{SumResult}
S(LMA-I)= 1.7 \pm 3.7 \quad ; \quad  S(LMA-II)=3.6 \pm 3.7
\end {equation}
show that both solutions are consistent with the sum rule.
The constraint (\ref{sum}), which is practically
unaffected if $r$ is varied within its assumed $\pm 20\%$ uncertainty,
will become relevant when more  data are available.

\subsection{Full spectral analysis}

\begin{figure}
\begin{center}
\epsfysize8truecm
\epsfbox{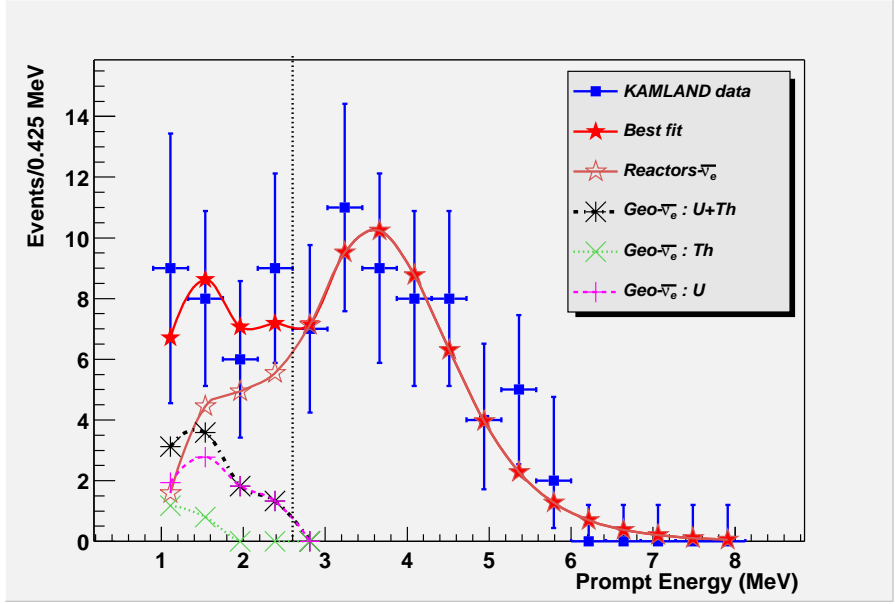}
\vspace{0.5cm}
\caption[a]{{\bf{Best fit to KamLAND data including geo-neutrinos 
with $[Th]/[U]=3.8$ (low-LMA)}.} The various contributions to the
sum spectrum are shown as derived by the fit, which gives 
 $N(U+Th)=9.9 \pm 6.2$. The dotted vertical line corresponds to $2.6\, MeV$.}
\label{fig1}
\end{center}
\end{figure}

\begin{figure}
\begin{center}
\epsfysize8truecm
\epsfbox{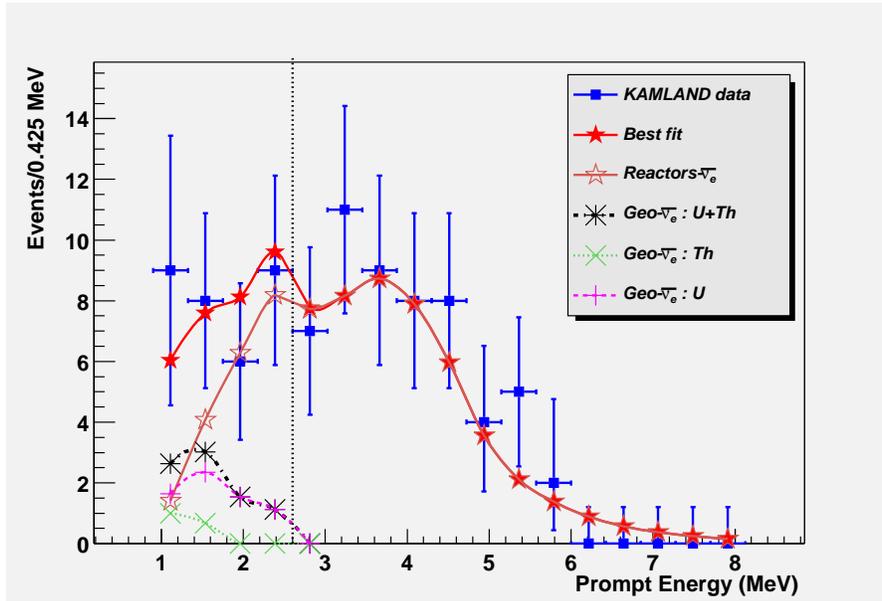}
\vspace{0.5cm}
\caption[a]{{\bf{Second best fit to KamLAND data including geo-neutrinos 
with $[Th]/[U]=3.8$ (high-LMA)}.} $N(U+Th)$ corresponds to $8.4 \pm 5.9$.}
\label{fig2}
\end{center}
\end{figure}

In addition to the algebraic approach described above, we performed a fit 
to the entire positron spectrum ($E_{vis}> 0.9 \, MeV$) including the geo-neutrino
contribution with $[Th]/[U]$ fixed at 3.8. The fitting function includes then the 
reactor fluxes of the 16 main contributing power plants, as well as the geo-neutrino 
spectrum of $^{238}U$ and $^{232}Th$ \cite{spectrum} convoluted with the KamLAND 
energy resolution of 7.5\%/$\sqrt{E(MeV)}$. Only the 5 first bins have a non-zero geo-neutrino 
contribution. According to \cite{kamland} we included 2.9 background events; since the exact 
background distribution has not been published, we added 2 of them into bin 1 and 0.3 into bin 
2, 3 and 4 respectively. We renormalised the no-oscillation spectrum to 86.9 events for 
$E_{vis} > 2.6 \, MeV$ in order to match the KamLAND integrated exposure. This leads to about
$122$ expected reactor events for $E_{vis} > 0.9 \, MeV$ in the absence of oscillations. 
It is worth noting
that in addition to the error of the overall normalisation (5.6 events), the lack of knowledge 
of the individual running time of the reactors adds another systematic error that we do not
take into account. 
The $\chi^2$ function is taken as in \cite{Nunokawa} which accounts for bins with low statistics. 
Fitting of the KamLAND spectrum leads to two main minima which we label 
low-LMA and high-LMA\footnote{For clarity, we refer to LMA-I and LMA-II  when discussing  results
of combined analysis of KamLAND ($E_{vis}>2.6\, MeV$) + Chooz + solar data,
we refer to low-LMA and high-LMA when considering the energy spectrum
of KamLAND.}. 
Both solutions remain stable when increasing the threshold to $E_{vis}> 2.6 \, MeV$; the best fit
 values do shift only slightly when including the additional geo-neutrino information.
The detailed results of the fit are given in Table \ref{table4} and displayed in Figs. \ref{fig1} 
and \ref{fig2}. 
The best fit is obtained for the low-LMA solution ($\chi^2 = 6.0/14$) and the corresponding 
geo-neutrino contribution $N(U+Th)$ is found to be $9.9 \pm 6.2$. 
The high-LMA solution is allowed at a one sigma level with $\Delta\chi^2 = 2.2$ and
$N(U+Th)$ is $8.4 \pm 5.9$. This result can be compared with 9 events obtained in \cite{kamland}.

We applied the algebraic method described in the previous section using the low-LMA and high-LMA 
solutions, and find $S_{low-LMA}=0.6 \pm 3.7$ and $S_{high-LMA}=4.8 \pm 5.0$.
This shows that the sum rule can be applied to check the consistency of the data and the geo-neutrino
predictions.

\begin{figure}
\begin{center}
\epsfysize18truecm
\epsfbox{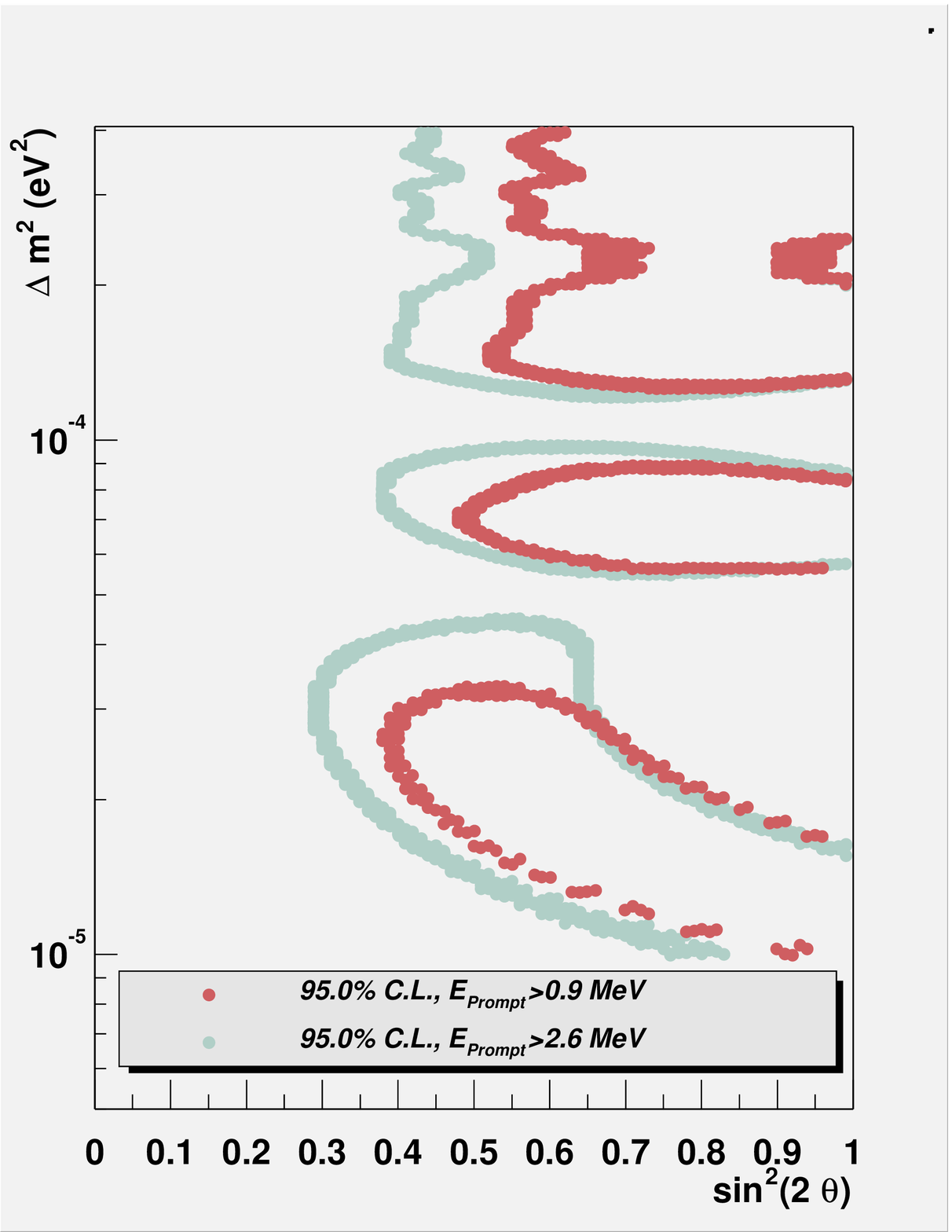}
\vspace{0.5cm}
\caption[a]{{\bf{Comparison of the 95\% C.L. allowed areas with and
without geo-neutrinos constraint}.} The light-gray region is allowed analyzing the
spectrum with an energy threshold of $2.6 \, MeV$, while the dark-gray region
is allowed including the full spectrum and constraining to $[Th]/[U]=3.8$.}
\label{fig3}
\end{center}
\end{figure}

Finally we compare the constraints on oscillation parameters obtained from the analysis with  
$E_{vis}> 2.6 \, MeV$ and $E_{vis}> 0.9 \, MeV$. In the latter case we include geo-neutrinos with
the ratio $[Th]/[U]$ fixed at 3.8 and background as described above. We calculate the 
values of $\chi^2$ in the 3-dimensional parameter space $[\Delta m^2, \sin^2 2\theta, N(U+Th)]$.
To obtain the 95\% C.L. for the subspace of interest, we then project the volumes which 
satisfy $\chi^2 - \chi^2_{min}< 5.99$ (joint estimation of 2 parameters) 
onto the $[\Delta m^2, \sin^2 2\theta]$ plane \cite{numrecipe}.
Fig.~\ref{fig3} shows the  95\% contour lines of the full spectral analysis 
together with the contours obtained for $E_{vis}> 2.6 \, MeV$.
The latter is in good agreement with \cite{kamland}, however we remark that our 90\% 
C.L. contour is closer to their 95\% C.L.. 

Even with the present limited statistics, the full spectral analysis further  
reduces the allowed oscillation parameter space compared to the analysis
with $E_{vis}> 2.6 \, MeV$. 
We expect that with an increased  statistics in the future, a full spectral analysis
including geo-neutrinos will provide a severe consistency check of the data and moreover can
help to break the degeneracy among the solutions, in particular, if a high-LMA solutions 
\cite{Schonert} were realized in nature.

\acknowledgements
It is a pleasure to acknowledge several interesting discussions with 
with  L.~Beccaluva, G.~Bellini, E.~Bellotti, C.~Bonadiman, C.~Broggini,
L.~Carmignani, T. Kirsten, E. Lisi, F.~Mantovani, G.~Ottonello and C.~Vaccaro.

\section*{Appendix: From flux to signal}

The number of events $N(X)$ from the decay chain of element $X=U,Th$ is:
\begin{equation}
\label{eqapp1}
N(X)=N_p\, t \int dE_{\bar{\nu}} \epsilon(E_{\bar{\nu}}) 
\sigma(E_{\bar{\nu}}) \varphi_X^{(arr)}(E_{\bar{\nu}})
%
%
\end{equation}
where $N_p$ is the number of free protons in the target, $t$ is the exposure time,
$\epsilon$ is the detection efficiency and 
$ \varphi_X^{(arr)}(E_{\bar{\nu}}) $
is the differential flux of antineutrinos arriving into the detector:
\begin{equation}
\label{eqapp2}
 \varphi_X^{(arr)}(E_{\bar{\nu}})  = \int _{V_\oplus} d^3r \frac{ \rho(\vec{r}) } {4\pi|\vec{R}-\vec{r}|^2} \,
				      \frac{C_X(r)n_X}{\tau_X m_X}\,
                                       f_X(E_{\bar{\nu}}) \,
                                      p(E_{\bar{\nu}},|\vec{R}-\vec{r}| )
\end{equation}
where $\rho$ is the density, $C_X$, $\tau_X$ and $m_X$ are the concentration, lifetime
and atomic mass of element $X$  and $n_X$ is the number of antineutrinos emitted per decay chain.
$f_X(E_{\bar{\nu}}) $ is the energy distribution of emitted antineutrinos, normalized to unity, and 
$p$ is the survival probability for $\bar{\nu}_e$  produced at $\vec{r}$ to reach the detector
at $\vec{R}$.

In view of the values of the oscillation length one can average the survival probability
over a short distance and bring out of the integral the term:
\begin{equation}
P=\langle p \rangle =1-\frac{1}{2} \sin^22\theta \quad.
\end{equation}

In this way we are left with:
\begin{equation}
\label{eqapp}
N(X)= N_p \, t \, P\, \int  dE_{\bar{\nu}}\, \epsilon(E_{\bar{\nu}})\, \sigma(E_{\bar{\nu}})\,  f_X(E_{\bar{\nu}}) \,
      \int _{V_\oplus} d^3r \frac{ \rho(\vec{r}) } {4\pi|\vec{R}-\vec{r}|^2} \,
				      \frac{C_X(r)n_X}{\tau_X m_X}\quad .
\end{equation}
The second integral is the produced flux of antineutrinos $\Phi(X)$ of eq.(\ref{eqflux}). Also
one can assume the detection efficiency as approximately constant over the small
($\approx 1\, MeV$) energy integration region. This leads to:
\begin{equation}
\label{eqapp4}
N(X)= N_p \, t \, P\, \epsilon(X)\Phi(X) \, \int  
      dE_{\bar{\nu}}  \sigma(E_{\bar{\nu}})f_X(E_{\bar{\nu}})
\end{equation}
This integral is easily computed from the cross section given
in Ref.~\cite{cross} and the spectrum from Ref.~\cite{spectrum}.

\begin{table}
\caption[aa]{{\bf{KamLAND results and theoretical predictions}}.
Events are estimated from \cite{noi}, for
$1.39 \cdot 10^{31}$ protons$\cdot$yr, 78.3\% efficiency and 
0.55 survival probability.}
\begin{tabular}{llll}
                         & $N(Th)$     &    $N(U)$  & $N(Th+U)$     \\      
KamLAND                  &  5          &          4 & 9          \\    
\hline
Chondritic               &  0.53       &       2.05 & 2.58    \\  
BSE                      &  0.62       &       2.45 & 3.07    \\  
Fully radiogenic         &  1.03       &       4.03 & 5.06     \\
\end{tabular}
\label{tabevents}
\end{table}

\begin{table}
\caption[ab]{{\bf{Estimated geo-neutrinos events }} for $10^{32}$ 
protons$\cdot$yr,
100\% efficiency and 0.55 survival probability.}
\begin{tabular}{lllll}
     Model   & [ref]  &         $N(Th)$       &    $N(U)$&            $N(Th+U)$  \\
\hline
Chondritic &\cite{noi}       &       4.8 &      18.9 &     23.7 \\
BSE    &\cite{noi}          &        5.7     &  22.5     & 28.2 \\
Fully Radiogenic  &\cite{noi}      & 9.5  &     37.0 &     46.5\\
\hline
from KamLAND data &\cite{kamland} & 45.9      & 36.7  &    82.6 \\
\hline
from &\cite{Chen}$^*$        &       5.5 &      21.1  &    26.6 \\
Ia &\cite{Ragh}   &                  8.6&       32.5  &    41.1\\
Ib &\cite{Ragh}                  &   5.7     &  21.6   &   27.3 \\
IIb & \cite{Ragh}  &                14    &     54    &    68  \\
\end{tabular}
\label{tabconf}
\footnotesize{
$^*$Values obtained from the fluxes of 
\cite{Chen} and eqs.(\ref{eqNU},\ref{eqNTh})}.
\end{table}

\begin{table}
\caption[abb]{ {\bf{Geo-events expected  for LMA-I and LMA-II and  the $[Th]/[U]$ constraint}.}
The best fit parameters ($\Delta m^2, \, \sin^2 2\theta$) are from
the combined analysis of KamLAND ($E_{vis}> 2.6 \, MeV$), solar and Chooz data from
\cite{Fogli}.
Survival probabilities $P_{a,b}$ of reactor antineutrinos from \cite{Fogli},
counts $C_{a,b}$, estimates of no-oscillation reactor events
$R_{a,b}(n.o.)$ and background $B_{a,b}$ from \cite{kamland}, 
Reactor events are estimated from $R=P\cdot R(n.o)$. The extracted geo-events are
$G=C-R-B$. Errors on the measured counts $C_{a,b}$ correspond to statistical
fluctuations.}
\begin{tabular}{lll}
Solution &  LMA-I  & LMA-II \\
\hline
$\Delta m^2 \, (10^{-5}eV^2)$  & 7.3  & 15.4 \\
$\sin^2 2\theta$                & 0.863  & 0.840\\
$P_a$                          & 0.65   & 0.60 \\
$P_b$  & 0.50 & 0.58 \\
$R_a(n.o.)$ & 10 & 10 \\
$R_b(n.o.)$ & 27 & 27 \\
$R_a$ & 6.5 & 6 \\
$R_b$ & 13.5 & 15.5 \\
$C_a$ & $17\pm 4.12$ & $17\pm 4.12$ \\
$C_b$ & $15\pm 3.87$ & $15\pm 3.87$ \\
$B_a$ & 3 & 3 \\
$B_b$ & 0 & 0 \\
$G_a=C_a-B_a-R_a$ & $7.5\pm 4.12$ & $8.0\pm 4.12$ \\
$G_b=C_b-B_b-R_b$ & $1.5\pm 3.87$ & $-0.5\pm 3.87$ \\
\hline
$\mathbf{N(U+Th)=G_a+G_b}$ & $\mathbf{9\pm 5.7}$ & $\mathbf{7.5\pm 5.7}$
\\
$\mathbf{S=0.4G_a-0.85G_b}$ & $\mathbf{1.7\pm 3.7}$ & $\mathbf{3.6\pm
3.7}$ \\
\end{tabular}
\label{table3}
\end{table}

\begin{table}
\caption[ac]{{\bf{Energy spectrum analysis with and without the
geo-neutrinos
constraint}.} Results from the full spectrum with
$[Th]/[U]=3.8$ constraint are compared  with those from
$E_{vis} > 2.6 \, MeV$.}
\begin{tabular}{lllll}
Range of fit  &\multicolumn{2}{c}{$E_{vis}>0.9 \, MeV$} &\multicolumn{2}{c}{$E_{vis}>2.6 \, MeV$} \\
Solution      &  low-LMA    &  high-LMA     & low-LMA    &  high-LMA  \\
\hline
$\Delta m^2 \, (10^{-5}\,eV^2)$  &  6.8  & 14.8  & 6.8          &  14.8      \\
$\sin^2 2\theta$                 &  0.91 & 0.84  & 0.92         &  0.78      \\
$N(U+Th)$                        &  9.9 $\pm$ 6.2  & 8.4 $\pm$ 5.9  &  -   &   - \\
\hline
$\chi^2$                         &   6.0  & 8.2  & 5.1   &  6.9 \\
Data points                      &   17   & 17   & 13    &  13  \\
d.o.f.                           &   14   & 14   & 11    &  11  \\
\end{tabular}
\label{table4}
\end{table}


\newpage

\begin{thebibliography}{99}

\bibitem{kamland}
K. Eguchi et al., KamLAND Collaboration, hep-ex/0212021.

\bibitem{noi}
G. Fiorentini F. Mantovani and B. Ricci, nucl-ex/0212008.

\bibitem{LB}
Landolt-B\"{o}rnstein, ``Numerical data and functional relationships
in science and technology'', New Series, Group IV vol. 3a, Springer-Verlag, Berlin 1993.
http://www.landolt-boernstein.com/; http://ik3frodo.fzk.de/beer/pub/PB\underline{ }198-203.pdf.

\bibitem{Anders}
E. Anders and N. Grevesse, Geoch. Cosmoch. Acta, 53 (1989) 197.

\bibitem{Ragh}
R.S. Raghavan et al. , Phys. Rev. Lett. 80 (1998) 635.

\bibitem{Chen}
C.G. Rothschild, M.C. Chen and F.P. Calaprice, nucl-ex/9710001.
Geophy. Research Lett. 25 (1998) 1083.

\bibitem{borexino}
 G. Alimonti et al., BOREXINO Collaboration, Science and Technology of BOREXINO,
Astropart.Phys. 16 (2002) 205-234.


\bibitem{Fogli}
G.L. Fogli et al., hep-ph/0112127.

\bibitem{Maltoni}
M. Maltoni, T Schwetz and J.W.F. Valle, hep-ph/012129.

\bibitem{Bah}
J.N. Bahcall, M.C. Gonzales-Garcia and C. Penya-Garay, hep-ph/0212147.


\bibitem{Nunokawa}
H. Nunokawa et al., hep-ph/0212202.

\bibitem{Aliani}
P. Aliani et al., hep-ph/0212212.

\bibitem{Holanda}
P.C. de Holanda and A.Yu. Smirnov, hep-ph/0212270.


\bibitem{Barger}
V. Barger and D. Marfatia, hep-ph/0212126.


\bibitem{Wede}
G.Harmann and K.H. Wedepohl, Geoch. et Cosm. Acta 57 (1993) 1761.

\bibitem{Ahrens}
T.A. Ahrens ed, ``Global Earth Physics: a handbook of physical constants'', 
American Geophysical Union, Washington, 1995.

\bibitem{cross}
P. Vogel and  J. Beacom, Phys. Rev. D 60 (1999) 053003.
C. Bemporad, G. Gratta, P. Vogel, Rev. Mod. Phys. 74 (2002) 297.

\bibitem{spectrum}
H.~Behrens  and J.~Janecke, ``Numerical tables for beta decay 
and electron capture'', Springer-Verlag, Berlin, 1969.

\bibitem{Schonert}
S. Sch\"{o}nert, T. Lasserre and L. Oberauer, hep-ex/0203013, Astrop. Phys. (2003) in print.


\bibitem{numrecipe} 
W.H. Press et al., ``Numerical recipes'', Cambridge University Press, p. 689 (2002).

\end{thebibliography}
\end{document}